%% file: main.tex
\documentclass[10pt, conference, letterpaper]{IEEEtran}

\IEEEoverridecommandlockouts

\input{commands}

\usepackage{balance}
\usepackage{cite}
\usepackage{amsmath,amssymb,amsfonts}
\usepackage{subcaption}
\usepackage[export]{adjustbox}
\usepackage{algorithmic}
\usepackage{graphicx}
\usepackage{textcomp}
\usepackage{xcolor}
\usepackage{multirow}
\usepackage{tabularx}
\usepackage{tikz}
\usepackage[hidelinks]{hyperref}

\usepackage{booktabs}

\usepackage{mwe}

\def\BibTeX{{\rm B\kern-.05em{\sc i\kern-.025em b}\kern-.08em
    T\kern-.1667em\lower.7ex\hbox{E}\kern-.125emX}}

\makeatletter
\newcommand{\linebreakand}{%
  \end{@IEEEauthorhalign}
  \hfill\mbox{}\par
  \mbox{}\hfill\begin{@IEEEauthorhalign}
}
\makeatother

\usetikzlibrary{
  arrows,
  arrows.meta,
  calc,
  colorbrewer,
  decorations.pathreplacing,
  external,
  fit,
  positioning,
  shapes.geometric,
}
\pgfdeclarelayer{bg}    
\pgfdeclarelayer{fg}    
\pgfsetlayers{bg,main,fg}
\tikzexternaldisable%

\begin{document}

\title{Understanding IoT Domain Names: Analysis and Classification Using Machine Learning
\thanks{This work has been supported by ANR and BMBF within the PIVOT project (https://pivot-project.info/)}
}

\author{\IEEEauthorblockN{Ibrahim Ayoub}
\IEEEauthorblockA{\textit{Afnic~--~Université Paris-Saclay}\\
Yvelines, France\\
ibrahim.ayoub@afnic.fr}
\and
\IEEEauthorblockN{Martine S. Lenders}
\IEEEauthorblockA{\textit{Technische Universität Dresden}\\
Dresden, Germany\\
martine.lenders@tu-dresden.de}
\and
\IEEEauthorblockN{Benoît Ampeau}
\IEEEauthorblockA{\textit{Afnic}\\
Yvelines, France\\
benoit.ampeau@afnic.fr}
\and
\IEEEauthorblockN{Sandoche Balakrichenan}
\IEEEauthorblockA{\textit{Afnic}\\
Yvelines, France\\
sandoche.balakrichenan@afnic.fr}

\linebreakand

\IEEEauthorblockN{Kinda Khawam}
\IEEEauthorblockA{\textit{Université de Versailles St-Quentin}\\
Versailles, France\\
kinda.khawam@uvsq.fr}
\and 
\IEEEauthorblockN{Thomas C. Schmidt}
\IEEEauthorblockA{\textit{HAW Hamburg}\\
Hamburg, Germany\\
t.schmidt@haw-hamburg.de}
\and
\IEEEauthorblockN{Matthias Wählisch}
\IEEEauthorblockA{\textit{Technische Universität Dresden~--~Barkhausen Institut}\\
Dresden, Germany\\
m.waehlisch@tu-dresden.de}
}

\maketitle

\begin{abstract}
    In this paper, we investigate the domain names of servers on the Internet that are accessed by IoT devices performing machine-to-machine communications. Using machine learning, we classify between them and domain names of servers contacted by other types of devices. By surveying past studies that used testbeds with real-world devices and using lists of top visited websites, we construct lists of domain names of both types of servers. We study the statistical properties of the domain name lists and train six machine learning models to perform the classification. The word embedding technique we use to get the real-value representation of the domain names is Word2vec. Among the models we train, Random Forest achieves the highest performance in classifying the domain names, yielding the highest accuracy, precision, recall, and $F_1$ score. Our work offers novel insights to IoT, potentially informing protocol design and aiding in network security and performance monitoring.

\end{abstract}


\input{intro}

\input{background}

\input{method_with_datasets}

\input{results}

\input{related-work}
\input{conclusion}

\label{lastpage}

\bibliographystyle{IEEEtran}
\bibliography{references}

\appendices

\input{appendix}
\balance%
\end{document}

%% file: commands.tex
\usepackage{soulpos}
\usepackage{pifont}

\usepackage{xspace}
\newcommand{\eg}{\textit{e.g.,}~}
\newcommand{\ie}{\textit{i.e.,}~}

\newcommand{\quotedliteral}[1]{{\tt \textquotesingle #1\textquotesingle}}

\makeatletter
\renewcommand{\paragraph}[1]{\vspace*{0.03in}\noindent{\bf #1.}\hspace{0.25ex \@plus1ex \@minus.2ex}}

\newcommand{\paragraphS}[1]{\vspace*{0.03in}\noindent{\bf #1}\hspace{0.25ex \@plus1ex \@minus.2ex}}

\DeclareRobustCommand
  \smallvdots{\vbox{\baselineskip1.5\p@ \lineskiplimit\z@
    \hbox{.}\hbox{.}\hbox{.}}}
\makeatother

%% file: intro.tex
\section{Introduction}%
\label{sec:introduction}

Domain name classification enables detecting both phishing and domain names generated by domain generation algorithms (DGAs)~\cite{nguyet2022, qiao2019, woodbridge2016}. Phishing domain names are used by malicious servers that pose as legitimate ones and lure users into providing sensitive information and credentials. On the other hand, DGAs run on malware-infected devices and generate domain names to help the infected devices contact the Command \& Control (C\&C) servers. The domain name classification techniques could also be applied in the Internet of Things (IoT) environments. IoT devices often need to communicate with servers on the Internet to which they connect using their domain names~\cite{saidi_2022}. 

In this paper, we study IoT from a different viewpoint by studying the domain names of the servers on the Internet that IoT devices interact with and classify between them and servers contacted by other types of devices. We are interested in IoT devices that perform strictly machine-to-machine (M2M) communications. The servers such devices contact might be IoT-specific backend servers to which they relay information, receive commands and updates, or other generic servers not exclusive to IoT.

We compile two lists of domain names. Using packet captures of real devices, initially filtering the traffic of IoT M2M devices, we construct a list of domain names of servers that are contacted by these devices. This list we call \emph{IoT~M2M~Names}. The remaining packet captures we use to construct a list of domain names of servers exclusively contacted by other types of devices, never by IoT M2M devices. This list we call \emph{Other~Names}.

For the rest of the paper, we will refer to IoT M2M Devices as \emph{IoT~M2M~Devices} and IoT devices that are not M2M, generic devices, and human users as \emph{Other~Devices}.

The end goal is to study the domain names of servers contacted by \emph{IoT~M2M~Devices} and classify between them and the domain names of servers that cater to \emph{Other~Devices}. 

Previous works~\cite{woodbridge2016, qiao2019, yang2021, rao2020} use lists of top visited websites as a negative class in domain name classification problems. We also use two top lists, namely Cisco Umbrella 1 Million~\cite{cisco} (hereafter referred to as \emph{Cisco}) and Tranco Top 1M~\cite{tranco,tranco_paper} (hereafter referred to as \emph{Tranco}) to evaluate the performance of the models with such lists and the validity of using them in IoT domain name classification.

First, we construct the two lists, \emph{IoT~M2M~Names} and \emph{Other~Names}.
We use the public datasets from 12 previous studies (See Appendix~\ref{AppendixA:datasets}). Filtering the traffic of \emph{IoT~M2M~Devices}, we use it to construct the \emph{IoT~M2M~Names} set. The remaining traffic, the traffic of \emph{Other~Devices}, is then used to construct the \emph{Other~Names} set. This is followed by a data pre-processing phase that includes several sanitization tests and a study of the statistical properties of the domain names. Finally, we train six machine learning models to classify between \emph{IoT~M2M~Names} and \emph{Others~Names}, \emph{Cisco}, \emph{Tranco} and a list comprising the three sets, and evaluate the performance of the machine learning models.

Our work studies IoT from a new perspective by studying the domain names of the servers contacted by \emph{IoT~M2M~Devices}. 

We study the composition and the statistical properties of these domain names. Such insight helps shed light on domain names of IoT servers used by \emph{IoT~M2M~Devices} as we present our findings about what the average domain name of such servers looks like and identify patterns and particularities if found. This could be helpful to model and generate M2M IoT domain names that align with the average domain name of that type—for instance, aiding protocol design such as name compression for constrained devices\cite{lenders-dns-cbor-06}. Furthermore, the machine learning models we trained successfully classified \emph{IoT~M2M~Names} and \emph{Others~Names}, and the performance evaluation we performed indicated the ability of the models to be generalized to unseen data. This capability could aid in detecting outliers in M2M IoT networks. For example, non-M2M traffic in networks consisting solely of M2M IoT devices could be detected.

The remainder of the paper is organized as follows. Section~\ref{sec:background} provides background information, while Section~\ref{sec:method} outlines the methodology employed. Section~\ref{sec:results} presents the results obtained and Section~\ref{sec:discussion} discusses the key takeaways. Finally, Section~\ref{sec:relatedwork} discusses related work, and, Section~\ref{sec:conclusion} concludes the work.

%% file: background.tex
\section{Background}%
\label{sec:background}

\subsection{IoT M2M Names and Other Names}

\emph{IoT~M2M~Devices} usually contact servers on the Internet to relay information about the physical world or from which they receive commands and firmware updates~\cite{saidi_2022}. These devices rely on domain names as an indirection mechanism to connect to IP~endpoints, simplifying maintenance. This allows, for example, a transparent change of server addresses, as only the mapping in the DNS would need to be changed. The \emph{IoT~M2M~Devices} are typically pre-configured with the domain names of the servers they might need to contact, and they obtain the addresses of these servers by resolving the domain names via DNS\@. Beyond address resolution, future IoT devices might also use DNS to identify the service bindings of these servers, \eg whether they use the TCP-based HTTP/2, the QUIC-based HTTP/3, or other services such as CoAP, using SVCB resource records~\cite{RFC-9460,draft-ietf-core-transport-indication}.

The domain names of such servers exhibit distinct construction characteristics influenced by various factors. An IoT backend server, for example, might have a name that correlates with its high-level function. For example, a collection of IP cameras might have a backend server whose domain name is \texttt{cam.example.com}. Moreover, large IoT backend service providers tend to adopt a naming convention for their servers, which follows the pattern below~\cite{saidi_2022}:
\begin{center}
\texttt{\small\textless{}subdomain\textgreater{}.\textless{}region\textgreater{}.\textless{}second-level-domain\textgreater{}},
\end{center}
where \texttt{\textless{}subdomain\textgreater{}} could be the name of the IoT service or the protocol name, \texttt{\textless{}region\textgreater{}} refers to the location of the server, and \texttt{\textless{}second-level-domain\textgreater{}} could be the second-level domain of the service provider or a name related to the IoT service. 

Last, being involved in M2M communications, some \emph{IoT~M2M~Names} may contain machine-friendly character sequences that do not prioritize legibility or memorability and are challenging for humans to comprehend.

Meanwhile, \emph{Other~Names}, \ie domain names of servers catering \emph{Other~Devices} do not follow the same patterns. Since servers with such domain names serve a wide range of devices and human users, the legibility and memorability of their domain names are prioritized.

In this work, we visualize the differences between the two types of domain names by studying their statistical properties. We then move to classify them using machine learning.

\subsection{Word2vec for Word Embedding}

Before inputting them into machine learning models, the domain names have to be processed to obtain a real-valued vector representation of them. There are several options to achieve this. One way is through Natural Language Processing (NLP). NLP methods aim to obtain the real-valued vector representation of the textual data. This includes, for example, character level embedding~\cite{durld2021, yang2021, qiao2019}, which gives each character a fixed-size real-valued vector representation. Another NLP method is the Term Frequency-Inverse Document Frequency (TF-IDF), which assigns an importance value to each text element based on its frequency of appearance~\cite{rao2020}. A different way of processing textual data is the extraction of hand-crafted features, which studies the text, tries to extract properties, and uses these properties as a real-valued vector representation of the text~\cite{rao2020, butnaru2021}.

The method we use in this paper to obtain the real-valued vector representation of the domain names is Word2vec. In the context of NLP, this is also called word embedding. Word2vec~\cite{word2vec_google, mikolov2013} is one of several word embedding techniques, and it uses a shallow neural network to convert each word to a vector of real numbers. Word2vec captures semantic relationships between words, and the resulting real-valued vectors depend on the context of each word within the text. Two possible architectures for Word2vec exist, Continuous Bag-of-Words (CBOW) and Skip-gram. CBOW estimates the vector representation of a target word based on the context (\ie surrounding words) of this word. The number of surrounding words considered within the context is specified by a \emph{window size}. Two words that appear regularly together in a text would have vector representations geometrically close to each other in space. Consequently, words that do not appear together in the text are assigned vector representations that are distant geometrically. For Skip-gram, the model predicts the words before and after a target word based on the \emph{window size}. Between CBOW and Skip-gram, we chose to go with CBOW as it is less expensive computationally and faster to train~\cite{mikolov2013}.

%% file: method_with_datasets.tex
\section{Method}%
\label{sec:method}

\begin{figure}
\centering
\resizebox{1\linewidth}{!}{%
  \input{method}%
}
\caption{Our method applied in this paper.}%
\label{fig:method}
\end{figure}
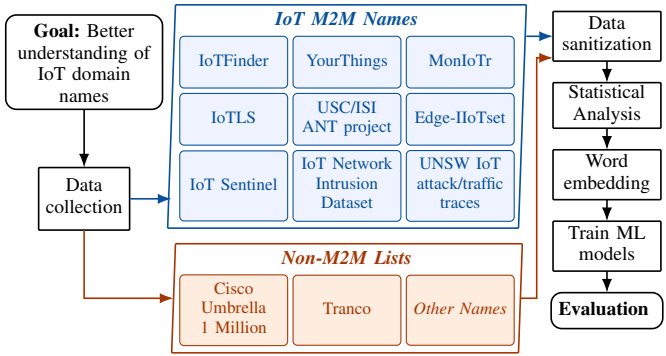

The objective of our study is to gain a better understanding of \emph{IoT~M2M~Names} and to evaluate the performance of common machine learning models in the classification between \emph{IoT~M2M~Names} and \emph{Other~Names}. We construct a dataset of domain names resolved by real \emph{IoT~M2M~Devices} and a dataset of domain names resolved by real \emph{Other~Devices}. In addition, we use two top lists, namely \emph{Cisco} and \emph{Tranco}. We sanitize the datasets to ensure that only valid domain names are used. We then familiarize ourselves with the datasets through a statistical analysis.

To train the machine learning models, we need a real-valued vector representation of the domain names, which we generate using Word2vec. Last, we train six machine learning models to classify between \emph{IoT~M2M~Names} and \emph{Other~Names}, \emph{Cisco}, and \emph{Tranco}. Figure~\ref{fig:method} summarizes our steps.

\subsection{Data Collection}%
\label{sec:data}
In our analysis, we use two types of domain name lists. The first type includes a list that contains domain names of servers contacted by \emph{IoT~M2M~Devices}, \ie \emph{IoT~M2M~Names}. The second type includes \emph{Other~Names}, a list of domain names of servers contacted by \emph{Other~Devices}, and two top lists, namely \emph{Cisco} and \emph{Tranco}. The following explains how we constructed the lists of domain names.

\subsubsection{\textbf{IoT~M2M~Names}}%
\label{subsec:iotdata}
The \emph{IoT~M2M~Names} list should contain domain names of servers on the Internet contacted by \emph{IoT~M2M~Devices}, \eg IoT backend servers that provide services to \emph{IoT~M2M~Devices} such as a server that saves the footage from IP cameras or servers from which \emph{IoT~M2M~Devices} receive commands and software updates. This list may also include domain names of servers that are not IoT-specific but are nevertheless contacted by \emph{IoT~M2M~Devices}.

We use 12 public datasets that have been gathered in prior work. The datasets we used are IoTFinder~\cite{iotfinder}, YourThings~\cite{yourthings}, MonIoTr~\cite{moniotr}, IoTLS~\cite{iotls}, three datasets from the USC/ISI ANT project~\cite{landerbootup2016, landerbootup2018, landertraces2020}, Edge-IIoTset~\cite{edge}, IoT Sentinel~\cite{sentinel}, IoT Network Intrusion Dataset~\cite{kang}, UNSW IoT traffic traces~\cite{unsw}, and UNSW IoT attack traces~\cite{unsw_attack}. See Appendix~\ref{AppendixA:datasets} for a summary of the 12 datasets we used. These datasets contain packet captures collected in testbeds that included real devices, of which are \emph{IoT~M2M~Devices}. Each is available as a set of PCAP files, including DNS messages sent from and received by the devices.

We filtered the PCAP files and extracted the unique DNS responses received by each device. Some datasets also contained captures from devices that were not strictly involved in M2M communication, such as desktop PCs, smartphones, gaming consoles, or smart TVs. We removed the captures of these devices but used them to construct \emph{Other~Names} (see below). Finally, we extracted the queried domain names from the resulting DNS responses. The resulting dataset, which we called \emph{IoT~M2M~Names}, contained 2551 unique domain names.

\subsubsection{\textbf{Other Names and Top-visited Websites Lists}}%
\label{subsec:noniot}
The \emph{Other Names} list should contain a list of domain names of servers used by devices not engaging in M2M communication or which are used by humans directly. We construct \emph{Other~Names} using traffic from the devices we excluded in the previous step. In addition, we use two top lists which are used as a negative class in domain name classification problems~\cite{woodbridge2016, qiao2019, yang2021, rao2020}. There are several lists of this kind, each with criteria for calculating popularity.

The three lists we use are:
\begin{itemize}
    \item \emph{\textbf{Other~Names:}} While preparing the \emph{IoT~M2M~Names} list, we filtered out the network traffic of devices that did not conform with our criteria for \emph{IoT~M2M~Devices}, \ie \emph{Other~Devices}. This allowed us to construct a dataset of real domain names of servers contacted by such devices. The dataset contained 6380 unique domain names. 
    \item \emph{\textbf{Cisco:}} The Cisco Umbrella 1 Million~\cite{cisco} is a daily published list of one million websites. Any domain name could be included in the list. The ranking of each domain name is based on the number of unique client IPs that visited it~\cite{cisco}.
    The list for our evaluation was gathered on September~21,~2023.
    \item \emph{\textbf{Tranco:}} Tranco~\cite{tranco} is a research-oriented list of one million domain names. The ranking of each domain name is based on its average rank over the past 30 days from four other popular domain name lists~\cite{tranco_paper}.
    The list for our evaluation was gathered on September~21,~2023, and thus covers the period from August~22 to September~20,~2023.
\end{itemize}

\subsection{Data Sanitization}%
\label{sec:datasanitization}

We conduct a data sanitization process, which includes several tests to ensure the validity of the domain names in each list. We perform the following tests: 

\paragraphS{Resolver Test. }%
\label{subsec:resolvable}
For each list of domain names, we try to resolve an \textit{A record} for every domain name. We ensure that domain names that exist but do not have the requested \textit{A record} are also counted as resolvable and account for empty non-terminal nodes.
The results of the resolver test are presented in Table~\ref{table:resolver}.

\begin{table}
  \centering
  \caption{Number of unique domain names after the resolver test.}%
  \label{table:resolver}
  \begin{tabular}{m{0.15\linewidth}m{0.2\linewidth}m{0.15\linewidth}m{0.1\linewidth}m{0.15\linewidth}}  
    \toprule
     Dataset & \emph{IoT~M2M~Names}      & \emph{Other~Names}     & \emph{Cisco}       & \emph{Tranco}      \\
    \midrule
    Resolvable            & 1\,417   & 4\,903     & 896\,093    & 970\,644    \\
    Unresolvable          & 1\,134   & 1\,477      & 103\,907    & 29\,356     \\
    \midrule
    Total                 & 2\,551   & 6\,380  & 1\,000\,000 & 1\,000\,000 \\
    \bottomrule
  \end{tabular}
\end{table}

\paragraphS{Syntax Check.}%
\label{subsec:syntax}
To ensure that all domain names in our lists respect the same syntax rules, we use the syntax checking used by Zonemaster~\cite{zonemaster}. The process starts with a normalization procedure that replaces all the dots with the regular full stop of Unicode \quotedliteral{\textbackslash u002E} (or \quotedliteral{.} as a character). The next step removes leading and trailing spaces. Next, a sequence of tests is conducted.

\begin{itemize}
  \item Check if the domain name starts with a dot,
  \item Check if the domain name has consecutive dots,
  \item Remove trailing dots if found,
  \item Check if any label in the domain name is longer than 63~characters,
  \item Check if the total length of the domain name is more than 253~characters,
  \item Check if the domain name has only one label,
  \item Check if any label starts or ends with a hyphen (\quotedliteral{-}), and, finally,
  \item Check if the domain name has double hyphen (\quotedliteral{--}) at positions 3 and 4 without it starting with \quotedliteral{xn}\footnote{\ie not an Internationalized Domain Name (IDN)}.
\end{itemize}

If one of the checks fails, the domain name is discarded. The results of the syntax check are presented in Table~\ref{table:syntax_check}.

\begin{table}
  \centering
  \caption{Number of unique domain names after the syntax check.}%
  \label{table:syntax_check}
  \begin{tabular}{m{0.15\linewidth}m{0.2\linewidth}m{0.15\linewidth}m{0.1\linewidth}m{0.15\linewidth}}  
    \toprule
     Dataset & \emph{IoT~M2M~Names}      & \emph{Other~Names}     & \emph{Cisco}       & \emph{Tranco}      \\
    \midrule
    Accepted              & 1\,415 & 4\,895   & 888\,297  & 970\,644 \\
    Discarded             & 2     & 8          & 7\,796    & 0        \\
    \midrule
    Total                 & 1\,417   & 4\,903     & 896\,093    & 970\,644    \\
    \bottomrule
  \end{tabular}
\end{table}

\paragraph{Remove commons}
Some domain names from the \emph{IoT~M2M~Names} might appear in \emph{Other~Names} or in the top lists. Therefore, we remove the common domain names between \emph{IoT~M2M~Names} and the other datasets from the other datasets. The resulting dataset sizes can be seen in Table~\ref{table:commons}.

\begin{table}
  \centering
  \caption{Number of unique domain names after removing the IoT domain names from the non-IoT datasets.}%
  \label{table:commons}
  \begin{tabular}{m{0.35\linewidth}m{0.15\linewidth}m{0.15\linewidth}m{0.15\linewidth}}
  
    \toprule
    Dataset                 & \emph{Other~Names}  & \emph{Cisco}     & \emph{Tranco}    \\
    \midrule
    Common with IoT Dataset & 979       & 940    & 14       \\
    Remaining               & 3\,916  & 887\,357  & 970\,630  \\
    \midrule
    Total                   & 4\,895  & 888\,297  & 970\,644  \\
    \bottomrule
  \end{tabular}
\end{table}

\paragraph{Final lists} After data sanitization, we obtain the final lists, which will be used in the following steps. The final lists can be seen in Table~\ref{table:finaldatasets}.

\begin{table}
  \centering
  \caption{Number of unique domain names in the final datasets.}%
  \label{table:finaldatasets}
  \begin{tabular}{m{0.26\linewidth}m{0.18\linewidth}m{0.15\linewidth}m{0.08\linewidth}m{0.08\linewidth}} 
    \toprule
     Dataset          & \emph{IoT~M2M~Names}      & \emph{Other~Names}     & \emph{Cisco}       & \emph{Tranco}      \\
    \midrule
    Domain Names [\#] & 1\,415  & 3\,916           & 887\,357 & 970\,630 \\
    \bottomrule
  \end{tabular}
\end{table}

\subsection{Statistical Study}%
\label{results-statistics}

\begin{figure}
\centering
\begin{subfigure}[b]{0.49\linewidth}%
  \includegraphics[width=\linewidth]{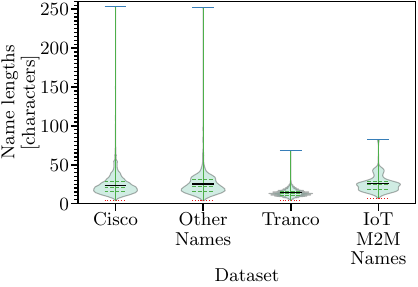}%
  \caption{Name lengths.}%
  \label{fig:statistics-names}%
\end{subfigure}
\begin{subfigure}[b]{0.46\linewidth}
    \centering
    \includegraphics[width=\linewidth]{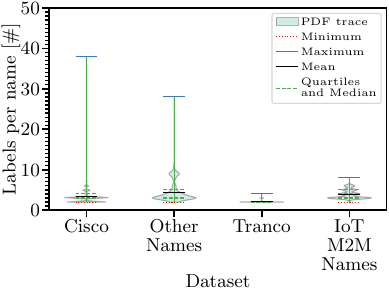}%
    \caption{Number of labels.}%
    \label{fig:statistics-labels}%
\end{subfigure}
\caption{Violin plots for name properties found for each domain name in our datasets.}%
\label{fig:statistics}
\end{figure}

We perform a statistical analysis of the domain name lengths and number of labels in each list, for which the results can be seen in the violin plots in Figure~\ref{fig:statistics}. Violin plots are similar to box plots, showing key statistical properties. However, they also estimate the probability density function (PDF) as a trace that forms the ``body'' of the ``violins'' around the properties.

The violin plots allow us to easily spot a similarity between domain names in \emph{IoT~M2M~Names}, \emph{Other~Names}, and \emph{Cisco} in terms of domain name length and number of labels per domain.
This is due to the way each dataset is constructed. The three datasets contain domain names as observed in the DNS requests and are, therefore, more representative.

\emph{Tranco}, on the other hand, is different from all the other lists we are using. The average Tranco domain name has fewer characters and labels than the average domain name from the other lists. Tranco mainly contains second-level domains in the form of \emph{domain.tld}, while \emph{IoT~M2M~Names}, \emph{Other~Names}, and \emph{Cisco} do not have that limitation.

\begin{figure*}
\centering
\begin{subfigure}[b]{0.2765\linewidth}%
    \includegraphics[width=\linewidth]{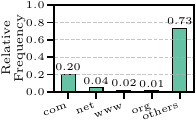}%
    \caption{Cisco.}%
    \label{fig:top-labels-cisco}%
\end{subfigure}%
\begin{subfigure}[b]{0.22\linewidth}
    \centering
    \includegraphics[width=\linewidth]{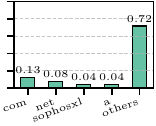}%
    \vskip-3.8pt
    \caption{Other Names.}%
    \label{fig:top-labels-excluded}%
\end{subfigure}%
\begin{subfigure}[b]{0.22\linewidth}%
    \centering
    \includegraphics[width=\linewidth]{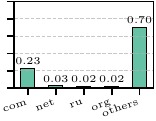}%
    \caption{Tranco.}%
    \label{fig:top-labels-tranco}%
\end{subfigure}%
\begin{subfigure}[b]{0.22\linewidth}%
    \centering
    \includegraphics[width=\linewidth]{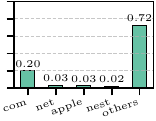}%
    \caption{IoT M2M Names.}%
    \label{fig:top-labels-iot}%
\end{subfigure}%
\caption{Top 50 labels in the datasets. The top 4 of those are shown separately, while the remaining 46 are summarized to ``others''.}%
\label{fig:top-labels}
\end{figure*}

In addition, we plot the relative frequencies of the top labels in each list in Figure~\ref{fig:top-labels}. The four lists have the same top two labels, which form at least 20\% of the total labels in each list. Moreover, the top labels across all lists are predominantly Top-Level Domains (TLDs), such as ``com'', ``net'', and ``org''. This implies that, when constructing an IoT M2M domain name, little emphasis is placed on selecting specialized IoT-only TLDs. Instead, the use of regular commercial TLDs seems a common practice. This indicates that the domain names of the different lists have, in general, similar TLDs and that the TLD of a domain name does not give a strong indication about the type of server that uses it as its domain name. We also see that the majority of labels---approximately 70\%--take less than or equal to 4\% each of the entire set.

The two studies above show that inspecting the length of a domain name in characters, its number of labels, or its TLD does not reveal information about whether this domain name is that of a server that caters to \emph{IoT~M2M~Devices} or \emph{Other~Devices}. In the context of our work, these values do not seem to be distinctive features to reveal the list the domain name belongs to. The only exception could be the domain names from \emph{Tranco}, which are second-level domains that tend to have limited number of characters.

\subsection{Word2vec: Real-Valued Vector Representation of Domain Names}

Word2vec expects prose text as input, \ie in the form of full documents with connected sentences and ideas where it can be used to capture the semantic relations. The challenge we face when using Word2vec with domain names is that these domain names do not form prose text. Instead, they are individual labels separated by periods.
As such, each domain name is treated as a sentence, and each label is treated as a word. For example, \texttt{iot.backend.org} contains three labels and is transformed to ``iot'', ``backend'', and ``org''. Another challenge is the limited size of domain names, which results in limited context and explains our choice of a \emph{window size} of 3. After the Word2vec algorithm is done, we obtain a real-valued vector representation of each label. The dimensions of each vector are set in advance. Before applying Word2vec, and to have a consistent dataset in terms of size for training the machine learning models, we pad the domain names by adding \quotedliteral{*} as a dummy label on the left of each domain name. The longest domain name in our lists has 38 labels, so we pad all the domain names to have 40 labels.

The parameters we used are as follows:
\begin{itemize}
  \item \textbf{Padding:} To each domain name, we added \quotedliteral{*} on the left. Each \quotedliteral{*} was treated as a dummy label (\ie a word), and they were added until all the domain names were of length 40 labels (words).
  \item \textbf{Word2vec:} We used CBOW (Continuous Bag-of-Words Model) with a \mbox{\emph{window size}} of 3.
  \item \textbf{Vectors:} Each word was represented by a vector $\in \mathbb{R}^{32}$.
\end{itemize}

\begin{figure}
  \centering%
  \begin{subfigure}{\linewidth}%
    \centering%
    \input{word2vec1}%
    \caption{Step 1: Generate Word2vec model as label to real-valued vector mapping.}%
    \label{fig:word2vec1}%
  \end{subfigure}\\
  \begin{subfigure}{\linewidth}%
    \centering%
    \input{word2vec2}%
    \caption{Step 2: Use Word2vec model to generate input for machine learning models from text corpus.}%
    \label{fig:word2vec2}%
  \end{subfigure}
\caption{Word embedding: After prepending \quotedliteral{*} to each domain name until it has 40 labels, Word2vec is used to generate a real-valued vector representation of $32 \times 40$ real numbers of each domain name.}%
\label{fig:word2vec}
\end{figure}
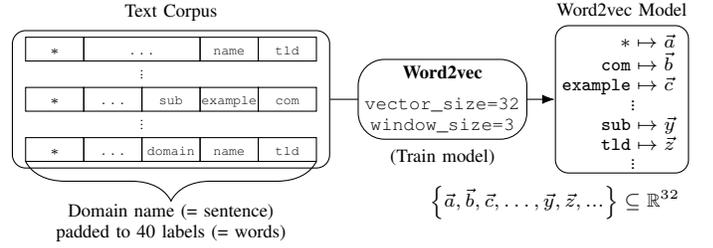
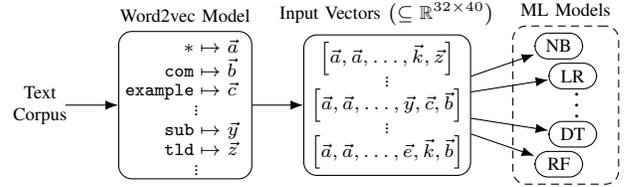

This vector representation can then be used to map each domain name to a $32 \times 40$ real-valued vector ($\in \mathbb{R}^{32 \times 40}$). The Word2vec process is depicted in Figure~\ref{fig:word2vec}.

%% file: method.tex
\begin{tikzpicture}[>=Latex]
  \tikzset{
    node/.style = {
      rectangle,
      draw,
      fill=none,
      font=\small\strut,
      align=center,
      node distance=1.1em and 1.1em,
      outer sep=0pt,
      rounded corners=0.5pt,
      line width=0.75pt,
    },
    start/.style = {
      node,
      rounded corners=5pt,
    },
    end/.style = {
      start,
      align=center,
    },
    process/.style = {
      node,
      inner ysep=0.5em,
    },
    decision/.style = {
      node,
      diamond,
      minimum width=5.5em,
      inner xsep=-2pt,
      inner ysep=-15pt,
      text depth = 0.3em,
    },
    input/.style = {
      node,
      trapezium,
      trapezium left angle=88,
      trapezium right angle=92,
      trapezium stretches=true,
      inner xsep=0.2em,
    },
    IoT/.style = {
      color=Blues-K,
    },
    IoT fill/.style = {
      fill=Blues-B,
    },
    non-IoT/.style = {
      color=Oranges-K,
    },
    non-IoT fill/.style = {
      fill=Oranges-B,
    },
    input-label/.style = {
      #1,
      font=\bfseries\small,
      outer sep=0pt,
      inner sep=0pt,
      align=center,
    },
    dataset/.style = {
      #1,
      node,
      font=\footnotesize\strut,
      rounded corners=2pt,
      node distance=0.25em,
      minimum height=2.4em,
      minimum width=5em,
      line width=0.5pt,
    },
    dataset IoT/.style = {
      dataset=IoT,
      IoT fill,
    },
    dataset non-IoT/.style = {
      dataset=non-IoT,
      non-IoT fill,
    },
    transition/.style = {
      -Latex,
      line width=0.75pt,
    },
  }

  \node [start, text width=6.6em] (Start) {{\bf Goal:} Better understanding of IoT domain names};
  \node [process, below=2.7em of Start] (Data collection) {Data\\collection};

  \node [dataset IoT, right=0.8em of Start] (IoTFinder) {IoTFinder};
  \node [dataset IoT, right=of IoTFinder] (YourThings) {YourThings};
  \node [dataset IoT, right=of YourThings] (MonIoTr) {MonIoTr};
  \node [input-label=IoT, above=0.5em of YourThings] (IoT dataset label) {\emph{IoT M2M Names}};
  \node [dataset IoT, below=of YourThings] (USC/ISI ANT) {USC/ISI\\ANT project\strut};
  \node [dataset IoT, left=of USC/ISI ANT] (IoTLS) {IoTLS};
  \node [dataset IoT, right=of USC/ISI ANT] (Edge-IIoTset) {Edge-IIoTset};
  \node [dataset IoT, below=of USC/ISI ANT, minimum height=3.25em] (IoT Network Intrusion) {IoT Network\\Intrusion\\Dataset};
  \node [dataset IoT, left=of IoT Network Intrusion, minimum height=3.25em] (IoT Sentinel) {IoT Sentinel};
  \node [dataset IoT, right=of IoT Network Intrusion, minimum height=3.25em] (UNSW IoT attack/traffic traces) {UNSW IoT\\attack/traffic\\traces};
  \node [input, IoT, fit={(IoT dataset label) (MonIoTr) (IoT Sentinel)}] (IoT dataset) {};

  \node [dataset non-IoT, below=2.5em of IoT Network Intrusion, minimum height=3.25em] (Tranco) {Tranco};
  \node [dataset non-IoT, left=of Tranco, minimum height=3.25em] (Cisco) {Cisco\\Umbrella\\1 Million};
  \node [dataset non-IoT, right=of Tranco, minimum height=3.25em] (Excluded Devices) {\emph{Other Names}};
  \node [input-label=non-IoT, above=0.5em of Tranco] (Non-IoT dataset label) {\emph{Non-M2M Lists}};
  \node [input, non-IoT, fit={(Non-IoT dataset label) (Cisco) (Excluded Devices)}] (Non-IoT dataset) {};
  \coordinate (first row center) at ($(Start.south west)!0.5!(Data collection.south east)$);

  \node [process, inner sep =2pt, minimum width=5em, right=1.8em of MonIoTr.north east] (Data sanitization) {Data\\sanitization};
  \node [process, minimum width=5em, inner sep =1pt, below=1em of Data sanitization] (Statistical Analysis) {Statistical\\Analysis};
  \node [process, inner sep =2pt, minimum width=5em, below=1em of Statistical Analysis] (Word embedding) {Word\\embedding};
  \node [process, inner sep =2pt, minimum width=5em, below=1em of Word embedding] (Train ML models) {Train ML\\models};

  \node [end, below=1em of Train ML models] (Evaluation) {%
   \textbf{Evaluation}
  };

  \draw [transition]                  (Start) -- (Data collection);
  \draw [transition, IoT]             (Data collection.east) -- (Data collection.east -| IoT dataset.west);
  \draw [transition, non-IoT]         ([yshift=-1em] Data collection) |- (Non-IoT dataset);
  \draw [transition, IoT]             (IoT dataset.top right corner |- Data sanitization.west) -- (Data sanitization);
  \draw [transition, non-IoT]         (Non-IoT dataset.east) -| ($(Non-IoT dataset.east)!0.5!(Data sanitization.west)$) |- ([yshift=-1em]Data sanitization.west);
  \draw [transition]                  (Data sanitization) -- (Statistical Analysis);
  \draw [transition]                  (Statistical Analysis) -- (Word embedding);
  \draw [transition]                  (Word embedding) -- (Train ML models);
  \draw [transition]                  (Train ML models) -- (Evaluation);
\end{tikzpicture}

%% file: word2vec1.tex
\tikzsetnextfilename{word2vec1}
\begin{tikzpicture}[>=Latex]
  \tikzset{
    domain-name/.style = {
      font=\tiny\tt,
      node distance=0.8em,
      outer sep=0pt,
    },
    input-word/.style = {
      rectangle,
      draw,
      fill=none,
      font=\tiny\tt,
      minimum height=0.9em,
      minimum width=2.2em,
      text height=0.6em,
      text depth=0.2em,
      inner sep=0pt,
      outer sep=0pt,
      node distance=0pt,
      align=center,
    },
    input-padding/.style = {
      input-word,
      font=\tiny$\ast$,
    },
    input-filling/.style = {
      input-word,
      fill=none,
      font=\tiny$\hdots$,
    },
    vdots/.style = {
      font=\tiny\smallvdots,
      inner sep=0pt,
      anchor=center,
      text height=0.55em,
      text depth=0.2em,
    },
    input/.style = {
      draw,
      font=\scriptsize,
      fill=none,
      rounded corners=0.5em,
      inner sep = 0.25em,
      minimum width = 4em,
      align = center,
      node distance = 3.5em,
    },
    operation/.style = {
      -Latex,
    },
    operation label/.style = {
      midway,
      above,
      font=\scriptsize,
    },
    word2vec/.style = {
      midway,
      align=center,
      font=\scriptsize,
      draw,
      fill=white,
      inner sep=0.25em,
      rounded corners=1em,
    },
    move/.style = {
      -Latex,
      densely dashed,
    }
  }

  \node [input-padding]                (domain 1 padding 2)  {};
  \node [input-filling, right=of domain 1 padding 2, minimum width=4.4em] (domain 1 filling 1)  {};
  \node [input-word, right=of domain 1 filling 1]                         (domain 1 name)       {name};
  \node [input-word, right=of domain 1 name]                              (domain 1 tld)        {tld};

  \node [input-padding, below=1.0em of domain 1 padding 2]              (domain 2 padding 2)  {};
  \node [input-filling, right=of domain 2 padding 2]                      (domain 2 filling 1)  {};
  \node [input-word, right=of domain 2 filling 1]                         (domain 2 sub)        {sub};
  \node [input-word, right=of domain 2 sub]                               (domain 2 example)    {example};
  \node [input-word, right=of domain 2 example]                           (domain 2 com)        {com};

  \node [input-padding, below=1.0em of domain 2 padding 2]                (domain 3 padding 2)  {};
  \node [input-filling, right=of domain 3 padding 2]                      (domain 3 filling 1)  {};
  \node [input-word, right=of domain 3 filling 1]                         (domain 3 domain)     {domain};
  \node [input-word, right=of domain 3 domain]                            (domain 3 name)       {name};
  \node [input-word, right=of domain 3 name]                              (domain 3 tld)        {tld};
  \node [vdots] at ($(domain 1 filling 1.south)!0.5!(domain 2 filling 1.north east)$) {};
  \node [vdots] at ($(domain 2 filling 1.south east)!0.5!(domain 3 filling 1.north east)$) {};
  \begin{pgfonlayer}{bg}
    \node [input, fit={(domain 1 padding 2) (domain 3 tld)}, label={\scriptsize Text Corpus}, inner xsep=0.5em] (text corpus) {};
  \end{pgfonlayer}
  \draw [decorate, decoration={brace, mirror, amplitude=1.5em}] (domain 3 padding 2.south west) -- (domain 3 tld.south east) node [midway, below=1.25em, align=center, font=\scriptsize] {Domain name (= sentence)\\padded to 40 labels (= words)};
  
  \node [input, label={\scriptsize Word2vec Model}, right=8.5em of text corpus] (word2vec model) {%
      $\begin{aligned}
        \mathtt{*}        & \mapsto \vec a \\[-0.5em]
        \mathtt{com}      & \mapsto \vec b \\[-0.5em]
        \mathtt{example}  & \mapsto \vec c \\[-0.5em]
                          & \text{\smallvdots} \\[-0.5em]
        \mathtt{sub}      & \mapsto \vec y \\[-0.5em]
        \mathtt{tld}      & \mapsto \vec z \\[-0.5em]
                          & \text{\smallvdots} \\[-0.5em]
      \end{aligned}$
  };

  \draw [operation] (text corpus) -- (word2vec model) node [word2vec, label=below:{\scriptsize (Train model)}] {{\scriptsize\bf Word2vec}\\[0.5em]{\tt vector\_size=32}\\{\tt window\_size=3}};

  \node [below=0.10em of word2vec model.south east, anchor=north east] {\scriptsize $\left\lbrace \vec a, \vec b, \vec c, \dots, \vec y, \vec z, ...\right\rbrace \subseteq \mathbb{R}^{32}$};
\end{tikzpicture}

%% file: word2vec2.tex
\tikzsetnextfilename{word2vec2}
\begin{tikzpicture}[>=Latex]
  \tikzset{
    input-word/.style = {
      rectangle,
      draw,
      fill=none,
      font=\tiny\tt,
      minimum height=0.9em,
      minimum width=2.2em,
      text height=0.6em,
      text depth=0.2em,
      inner sep=0pt,
      outer sep=0pt,
      node distance=1em and 0pt,
      align=center,
    },
    input-padding/.style = {
      input-word,
      font=\tiny$\ast$,
    },
    input-filling/.style = {
      input-word,
      fill=none,
      font=\tiny$\hdots$,
    },
    vdots/.style = {
      font=\tiny\smallvdots,
      inner sep=0pt,
      anchor=center,
      text height=0.55em,
      text depth=0.2em,
    },
    input/.style = {
      draw,
      font=\scriptsize,
      fill=none,
      rounded corners=0.5em,
      inner sep = 0.25em,
      minimum width = 4em,
      align = center,
      node distance = 3.5em,
    },
    operation/.style = {
      -Latex,
    },
    operation label/.style = {
      midway,
      above,
      font=\footnotesize,
    },
    word2vec/.style = {
      midway,
      font=Word2vec,
      draw,
      fill=white,
      inner sep=1em,
      rounded corners=1em,
    },
    move/.style = {
      -Latex,
      densely dashed,
    }
  }

    \node (anchor) {};
    \node[font=\scriptsize, align=center] at ([xshift=-0.25em]anchor.west) (text corpus) {Text\\Corpus};
  
  \node [input, label={\scriptsize Input Vectors \scriptsize$\left(\subseteq \mathbb{R}^{32 \times 40}\right)$}, right=9em of anchor] (vectors) {%
    $\left\lbrack \vec{a}, \vec{a}, \hdots, \vec{k}, \vec{z} \right\rbrack$\\[0.1em]
    \smallvdots\\[0.1em]
    $\left\lbrack \vec{a}, \vec{a}, \hdots, \vec{y}, \vec{c}, \vec{b} \right\rbrack$\\[0.1em]
    \smallvdots\\[0.1em]
    $\left\lbrack \vec{a}, \vec{a}, \hdots, \vec{e}, \vec{k}, \vec{b} \right\rbrack$
  };

  \node [input, fill=white, label={\scriptsize Word2vec Model}, right=2em of anchor] (word2vec model) {%
      $\begin{aligned}
        \mathtt{*}        & \mapsto \vec a \\[-0.5em]
        \mathtt{com}      & \mapsto \vec b \\[-0.5em]
        \mathtt{example}  & \mapsto \vec c \\[-0.5em]
                          & \text{\smallvdots} \\[-0.5em]
        \mathtt{sub}      & \mapsto \vec y \\[-0.5em]
        \mathtt{tld}      & \mapsto \vec z \\[-0.5em]
                          & \text{\smallvdots} \\[-0.5em]
      \end{aligned}$
  };

  \draw [operation] (anchor) -- (word2vec model);
  \draw [operation] (word2vec model) -- (vectors);

  \node [input, minimum width=1.8em] (NB) at ($(vectors.east) + ( 34:4em)$) {NB};
  \node [input, minimum width=1.8em] (LR) at ($(vectors.east) + ( 16:4em)$) {LR};
  \node [yshift=0.3em] at                  ($(vectors.east) + (  0:4em)$) {$\vdots$};
  \node [input, minimum width=1.8em] (DT) at ($(vectors.east) + (-16:4em)$) {DT};
  \node [input, minimum width=1.8em] (RF) at ($(vectors.east) + (-34:4em)$) {RF};

  \draw [operation] (vectors) -- (NB);
  \draw [operation] (vectors) -- (LR);
  \draw [operation] (vectors) -- (DT);
  \draw [operation] (vectors) -- (RF);

  \begin{pgfonlayer}{bg}
    \node [input, densely dashed, fit={(NB) (LR) (DT) (RF)}, label={\scriptsize ML Models}] {};
  \end{pgfonlayer}
\end{tikzpicture}

%% file: results.tex
\section{Results: Domain Name Classification}%
\label{sec:results}

We train six machine learning models to classify \emph{IoT~M2M~Names}, and \emph{Other~Names}, \emph{Cisco} and \emph{Tranco} domain names. We use the following models and refer to them using the acronyms in parentheses:
Na\"{\i}ve~Bayes~(NB), Logistic~Regression~(LR), K-Nearest~Neighbors~(KNN), Support~Vector~Machine~(SVM), Decision~Tree~(DT), and Random~Forest~(RF).

After the sanitization process, \emph{IoT~M2M~Names} contains 1415 domain names. From \emph{Other~Names}, \emph{Cisco}, and \emph{Tranco}, we then pick 1415 domain names individually. We also create an additional list of 1415 domain names by uniformly sampling a \emph{Mix} of \emph{Other~Names}, \emph{Cisco}, and \emph{Tranco}.

We select the 1415 domain names from \emph{Other~Names}, \emph{Cisco}, and \emph{Tranco} in two ways:
\begin{itemize}
  \item We take the top 1415 domain names or
  \item randomly choose them, uniformly distributed, from the whole list.
\end{itemize}

After selecting 1415 domain names from each list, the domain names are labeled accordingly, and a combined list is constructed. The combined list is then processed via Word2vec to obtain the real-valued vector representation of each domain name. These real-valued vectors are used to train the machine learning models. The models are trained as binary classifiers between two classes, namely \emph{IoT~M2M~Names} and \emph{Other} domain names where \emph{Other} domain names belong to either one of \emph{Other~Names}, \emph{Cisco}, or \emph{Tranco}, or a \emph{Mix} of them.
To evaluate the performance of each of the models, we calculate the resulting accuracy, precision, recall, and the $F_1$ score in subsection \ref{subsec:performance}. Moreover, we perform in subsection \ref{subsec:crossvalidation} cross-validation to assess the robustness of the models and their ability to generalize to unseen data. Lastly, we perform in subsection \ref{subsec:Ablation} an ablation test to analyze the impact of the different labels of the domain names on the performance of the models.

\subsection{Performance Evaluation}%
\label{subsec:performance}
In this section, we present our results after training several machine learning models to classify between \emph{IoT~M2M~Names} and \emph{Other} domain names. In each scenario, each list was processed with Word2vec to obtain the real-valued vector representation of each domain name of size $32 \times 40$. We used an 80-20 train-test split.

We train the machine learning models NB, LR, KNN, SVM, DT, and RF\@. For each model, we calculate four parameters: Accuracy, precision, recall, and the $F_{1}$ score. We first calculate the confusion matrix to identify true positives ($\mathrm{TP}$), true negatives ($\mathrm{TN}$), false positives ($\mathrm{FP}$), and false negatives ($\mathrm{FN}$). The values for our parameters are then produced using the following formulas:
\begin{align}
  \mathrm{Accuracy}   &= \frac{\mathrm{TP} + \mathrm{TN}}{\mathrm{TP} + \mathrm{TN} + \mathrm{FP} + \mathrm{FN}} \label{eq:accuracy}\\
  \mathrm{Precision}  &= \frac{\mathrm{TP}}{\mathrm{TP} + \mathrm{FP}} \label{eq:precision} \\
  \mathrm{Recall}     &= \frac{\mathrm{TP}}{\mathrm{TP} + \mathrm{FN}} \label{eq:recall}\\
  F_1                 &= \frac{2}{\frac{1}{\mathrm{Precision}} + \frac{1}{\mathrm{Recall}}} = \frac{\mathrm{TP}}{\mathrm{TP} + \frac{\mathrm{FN} + \mathrm{FP}}{2}} \label{eq:f1}
\end{align}

Accuracy measures the ratio of positive predictions ($\mathrm{TP}$ and $\mathrm{TN}$) to all the predictions made by the model, see Equation~\ref{eq:accuracy}. Precision measures the ratio of true positive predictions ($\mathrm{TP}$) to all the positive predictions ($\mathrm{TP}$ and $\mathrm{FP}$) made by the model, see Equation~\ref{eq:precision}. Recall measures the ratio of true positive predictions ($\mathrm{TP}$) to all the actual positive instances in the dataset ($\mathrm{TP}$ and $\mathrm{FN}$), see Equation~\ref{eq:recall}. Finally, the $F_1$ score is the harmonic mean of precision and recall, see Equation~\ref{eq:f1}.

\begin{figure}[t]%
  \centering%
  \begin{subfigure}{\linewidth}%
    \centering%
    \adjustbox{clip, trim={0 18.1mm 0 0}}{
      \includegraphics[width=0.9\linewidth]{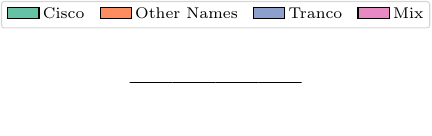}%
    }
  \end{subfigure}\\%
  \begin{subfigure}{0.5\linewidth}%
    \centering%
    \includegraphics[width=\linewidth]{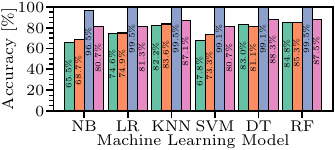}%
    \caption{Accuracy.}%
    \label{fig:results-top:accuracy}
  \end{subfigure}%
  \begin{subfigure}{0.5\linewidth}%
    \centering%
    \includegraphics[width=\linewidth]{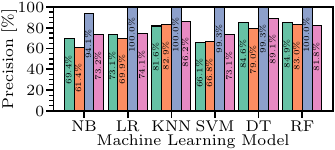}%
    \caption{Precision.}%
    \label{fig:results-top:precision}
  \end{subfigure}\\%
  \begin{subfigure}{0.5\linewidth}%
    \centering%
    \includegraphics[width=\linewidth]{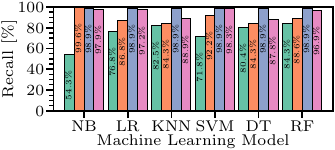}%
    \caption{Recall.}%
    \label{fig:results-top:recall}
  \end{subfigure}%
  \begin{subfigure}{0.5\linewidth}%
    \centering%
    \includegraphics[width=\linewidth]{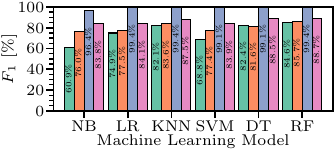}%
    \caption{$F_1$ score.}%
    \label{fig:results-top:f1}
  \end{subfigure}\\%
  \caption{Accuracy, precision, recall, and $F_{1}$ score of each ML model for the top 1415 domain names from \emph{Other~Names}, \emph{Cisco}, and \emph{Tranco}, plus a uniformly sampled \emph{Mix} of 1415 domain names from the three lists, each vs.\ the 1415 domain names from \emph{IoT~M2M~Names}.}%
  \label{fig:results-top}
\end{figure}

\begin{figure}[t]%
  \centering%
  \begin{subfigure}{\linewidth}%
    \centering%
    \adjustbox{clip, trim={0 18.1mm 0 0}}{
      \includegraphics[width=0.9\linewidth]{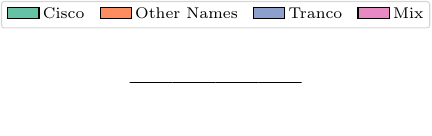}%
    }
  \end{subfigure}\\%
  \begin{subfigure}{0.5\linewidth}%
    \centering%
    \includegraphics[width=\linewidth]{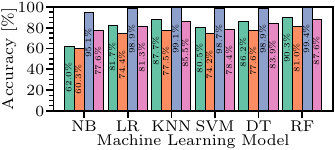}%
    \caption{Accuracy.}%
    \label{fig:results-random:accuracy}
  \end{subfigure}%
  \begin{subfigure}{0.5\linewidth}%
    \centering%
    \includegraphics[width=\linewidth]{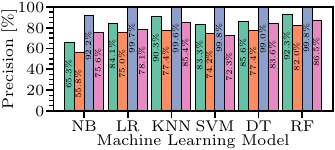}%
    \caption{Precision.}%
    \label{fig:results-random:precision}
  \end{subfigure}\\%
  \begin{subfigure}{0.5\linewidth}%
    \centering%
    \includegraphics[width=\linewidth]{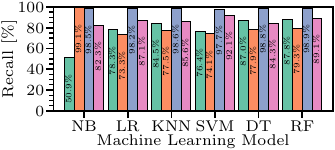}%
    \caption{Recall.}%
    \label{fig:results-random:recall}
  \end{subfigure}%
  \begin{subfigure}{0.5\linewidth}%
    \centering%
    \includegraphics[width=\linewidth]{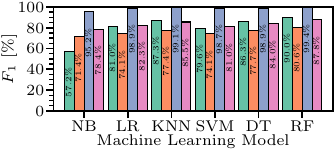}%
    \caption{$F_1$ score.}%
    \label{fig:results-random:f1}
  \end{subfigure}\\%
  \caption{Average accuracy, precision, recall and $F_{1}$ score of each ML model for 20 random picks of 1415 domain names from \emph{Other~Names}, \emph{Cisco}, and \emph{Tranco}, plus a uniformly sampled \emph{Mix} of 1415 domain names from the three lists, each vs.\ the 1415 IoT domain names.}%
  \label{fig:results-random}
\end{figure}

The results for using the top 1415 domain names can be seen in Figure~\ref{fig:results-top}. For the random selection of domain names, see Figure~\ref{fig:results-random}. 

\begin{figure}
  \centering
  \includegraphics{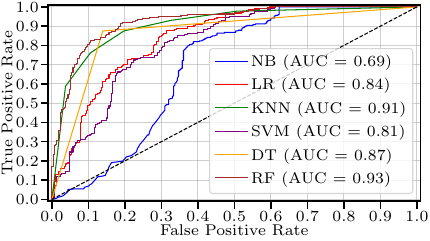}
  \caption{Receiver Operation Characteristic (ROC) Curves for the top 1415 domain names from \emph{Other~Names}. The Area Under the Curve (AUC) is provided in the legend.}%
  \label{fig:roc-excluded-top}
\end{figure}

\subsubsection{\textbf{Results when using top domain names from Other~Names, Cisco, and Tranco}}
The results of training the models using the top 1415 domain names from \emph{Other~Names}, \emph{Cisco}, and \emph{Tranco} are presented in Figure~\ref{fig:results-top}. Each graph represents one of the four parameters obtained by the different models.
The six models we trained exhibited the strongest performance when \emph{Tranco} was used. The lowest performing model when \emph{Tranco} was used was NB, but it still achieved values greater than $94\%$ for the four parameters, while the rest achieved values between $98\%$ and $100\%$ for the four parameters. The lowest-performing model, regardless if \emph{Other~Names}, \emph{Cisco}, or \emph{Tranco} was used, is NB\@. NB achieved values below $70\%$ for the four parameters when \emph{Cisco} was used and low accuracy, precision, and $F_1$ scores when \emph{Other~Names} and \emph{Mix} were used. NB, however, achieved high recall values of values $> 97.9\%$ when \emph{Other~Names}, \emph{Tranco} and \emph{Mix} were used. The best-performing overall model with all the lists is RF\@. RF achieved close to $99\%$ for the four parameters when \emph{Tranco} was used. Moreover, RF achieved slightly lower values for the four parameters for the rest of the lists, achieving values between $87\%$ and $92\%$ for the four parameters with \emph{Cisco} and \emph{Mix}. Even lower values were achieved for the four parameters when \emph{Other~Names} was used with values ranging between $79.3\%$ and $82\%$. The performance of the models is also visualized in Figure~\ref{fig:roc-excluded-top}, which shows the Receiver Operating Characteristic (ROC) curves plotted for every model when \emph{Other~Names} is used. ROC curves show the performance of the models at different classification thresholds. A comparison can be done between the performances of the models by comparing the Area Under the Curve (AUC) of each one. In our case, the ROC curves in Figure~\ref{fig:roc-excluded-top} further show the superiority of RF compared to the other models where its AUC~=~0.93. NB, as expected from the previous measurements, has the lowest AUC of 0.69 and, therefore, has the lowest performance between the six models.

\subsubsection{\textbf{Results when using random domain names from Other~Names, Cisco, and Tranco}}
The results of training the models when using random 1415 domain names from \emph{Other~Names}, \emph{Cisco}, and \emph{Tranco} are presented in Figure~\ref{fig:results-random}. Each graph represents one of the four parameters obtained by the different models. We trained the six models by randomly choosing 1415 domain names from each list to generalize our results further. This is particularly interesting when using \emph{Cisco} and \emph{Tranco}, each containing close to 1 million domain names. For each list, 100 random picks of 1415 domain names were made, and the results presented are the average of each of the four parameters over the 100 random picks.

The results over the 100 random picks are consistent with the previous results obtained when using the top domain names of \emph{Other~Names}, \emph{Cisco}, and \emph{Tranco}. The models performed best when \emph{Tranco} was used, with NB having the lowest performance and the rest of the models achieving values $> 98\%$ for the four parameters. The lowest performing model, regardless of the list used is still NB, particularly when \emph{Cisco} is used as a non-IoT dataset where it achieved values in the order of $60\%$ for the four parameters. NB, however, achieved high recall values of values $> 82.3\%$ when the lists other than \emph{Cisco} were used. The best-performing model, regardless of the list used is still RF\@. RF achieved values in the order of $99\%$ for the four parameters when \emph{Tranco} was used and had the lowest performance when \emph{Other~Names} was used with values in the order of $80\%$ for the four parameters.

\subsection{Cross Validation}%
\begin{figure}[t]%
  \centering%
  \begin{subfigure}{\linewidth}%
    \centering%
    \adjustbox{clip, trim={0 18.1mm 0 0}}{
      \includegraphics[width=0.9\linewidth]{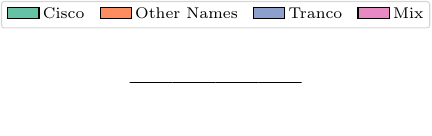}%
    }
  \end{subfigure}\\%
  \begin{subfigure}{0.5\linewidth}%
    \centering%
    \includegraphics[width=\linewidth]{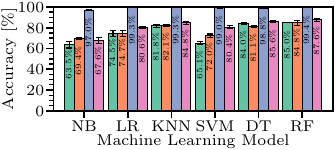}%
    \caption{Accuracy.}%
    \label{fig:crossvalidation:accuracy}
  \end{subfigure}%
  \begin{subfigure}{0.5\linewidth}%
    \centering%
    \includegraphics[width=\linewidth]{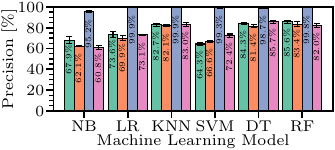}%
    \caption{Precision.}%
    \label{fig:crossvalidation:precision}
  \end{subfigure}\\%
  \begin{subfigure}{0.5\linewidth}%
    \centering%
    \includegraphics[width=\linewidth]{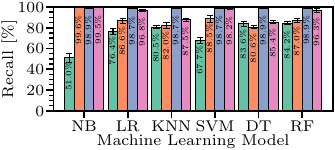}%
    \caption{Recall.}%
    \label{fig:crossvalidation:recall}
  \end{subfigure}%
  \begin{subfigure}{0.5\linewidth}%
    \centering%
    \includegraphics[width=\linewidth]{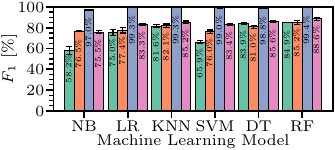}%
    \caption{$F_1$ score.}%
    \label{fig:crossvalidation:f1}
  \end{subfigure}\\%
  \caption{Mean (colored bars) and standard deviation (error bars) of accuracy, precision, recall and $F_{1}$ score of the ML models over five folds for the top 1415 domain names from \emph{Other~Names}, \emph{Cisco}, and \emph{Tranco}, plus a uniformly sampled \emph{Mix} of 1415 domain names from the three lists, vs.\ the 1415 IoT domain names.}%
  \label{fig:crossvalidation}
\end{figure}

\label{subsec:crossvalidation}
We use cross-validation to assess the robustness of our models and their ability to generalize to unseen data. When assessing a model using cross-validation, the dataset containing all the classes is divided into $K$ folds or subsets, and the model is trained $K$ times. During every training instance, one of the $K$ folds is used as a testing dataset, while the remaining $K-1$ are used for training. We use Stratified $K$-fold cross-validation to ensure that the distribution of classes in the folds is similar to their distribution in the original dataset. Given the size of the \emph{IoT~M2M~Names} list, we used $K$ = 5 to ensure that each fold contains enough entries to provide a reliable estimate of performance. $K$ = 5 allows each model to be trained five times. From \emph{Other~Names}, \emph{Cisco}, and \emph{Tranco}, we choose the top 1415 domain names, which are added to the 1415 domain names of \emph{IoT~M2M~Names}. We show the results in Figure~\ref{fig:crossvalidation} as averages and standard deviation values of the evaluation parameters over the five folds. 

The colored bars in Figure~\ref{fig:crossvalidation} represent the mean of the four parameters over the five folds, and the error bars at the top of each colored bar represent the standard deviation. We notice that the means of the four parameters over the five folds are consistent with the results from the performance evaluation we performed in Section~\ref{subsec:performance} while having a low standard deviation which indicates that the models are stable across the folds and that they are likely to generalize well to unseen data.

\subsection{Ablation Test}%
\label{subsec:Ablation}

\begin{figure}
\centering
\begin{subfigure}{0.24\textwidth}%
    \includegraphics[height=3cm, width=\linewidth]{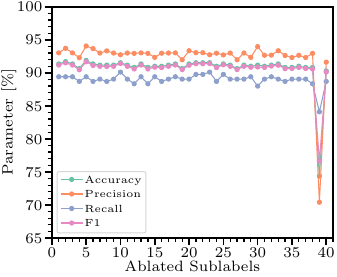}%
    \caption{Cisco.}%
    \label{fig:ablation-cisco-rf}%
\end{subfigure}%
\begin{subfigure}{0.24\textwidth}%
    \includegraphics[height=3cm, width=\linewidth]{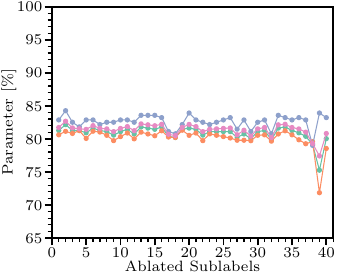}%
    \caption{Excluded Device.}%
    \label{fig:ablation-excluded-rf}%
\end{subfigure}%
\caption{Ablation Test with Random Forest (RF).}%
\label{fig:ablation}
\end{figure}

An ablation test includes removing elements from the ML model or suppressing a subset of the features to study the effect they might have on the performance. We perform the ablation test by removing one label at a time by replacing the 32-dimensional vector representing the label with zeros. We perform 40 training and testing sessions, ablating one label in both the training and the testing datasets every time, training, and finally evaluating the models. The results for RF when used with \emph{Other~Names} and \emph{Cisco} are presented in Figures~\ref{fig:ablation-cisco-rf}~and~\ref{fig:ablation-excluded-rf}.

The stable performance observed when ablating the dummy labels (\quotedliteral{*}) demonstrate that the padding we added to the left of each domain name held no information and did not alter the performance of the models. The second-to-last label, which is the second-level domain (label 39), seems to have the highest impact on the performance as the values drastically dropped in both figures. For example, the accuracy and precision in Figure~\ref{fig:ablation-excluded-rf} dropped by around 15 and 25 percent, respectively. When the last label---the TLD of the domain name (label 40)---was ablated, however, values of the parameters did not experience the same drastic decrease, and the effect of ablating label 40 seemed equivalent to ablating the dummy labels. This shows that the second-level domain of a domain name is the most indicative of its class, while other labels including the TLD do not provide information about it. This is consistent with the results obtained in Figure~\ref{fig:top-labels}, which showed that the TLDs are common between all the lists of domain names. Therefore, they are not distinctive about the class of the domain name.

\section{Discussion}%
\label{sec:discussion}

\paragraph{The size of \emph{IoT M2M Names}}
 
Despite the large amounts of raw data we started with, the size of \emph{IoT~M2M~Names} remained relatively comparatively modest after removing duplicates and sanitizing the data. This is primarily due to the scope of our study which covers IoT devices that engage in M2M communications. Such devices exhibit limitations in their functionalities compared to IoT devices that are not strictly M2M which explains the low number of servers on the Internet these devices contact. Hence, we noticed a low number of frequently contacted servers in contrast to numerous servers that are less regularly or rarely contacted. 

\paragraph{Usage and limitations of top-lists}
We used 2 known top-lists, namely \emph{Cisco} and \emph{Tranco}, to evaluate the validity of using such lists as negative class in contexts similar to ours. We noticed that \emph{Tranco} is the least valuable between the two as the majority of domains in it are second-level domains which does not reflect how domain names actually appear in DNS traffic. \emph{Cisco}, on the other hand, is suitable as the entries in it are not limited to second-level domains and are included in the list as seen in DNS traffic. \emph{Cisco} also resembles the domain names in \emph{IoT~M2M~Names} and \emph{Other~Names} statistically. The difference between \emph{Cisco} and \emph{Tranco} was most visible when training and testing the machine learning models. \emph{Tranco} is easily distinguishable and so the models almost achieved perfect scores with it. The performance was different with \emph{Cisco} which achieved a lower performance, but one which is comparable to when \emph{Other~Names}, the list that contains real domain names from real devices, was used. \emph{Cisco} seems to be the better option to be used as a negative class in domain name classification problems with similar contexts.

\paragraph{Better sources of data}
The domain names we used in this work, namely \emph{IoT~M2M~Names} and \emph{Other~Names}, were extracted from packet captures of testbeds that had real devices. A better data source would undoubtedly require larger, more diverse, testbeds that have more devices conforming with our criteria for \emph{IoT~M2M~Devices}. On the other hand, changing the scope to include devices other than strictly M2M ones, \eg smart TVs, would certainly enlarge and diversify the list of domain names.

%% file: related-work.tex
\section{Related Work}\label{sec:relatedwork}

The Identification of IoT devices is a popular topic in IoT research, primarily relying on machine learning techniques to accurately distinguish various features of IoT devices. The IoTFinder study~\cite{iotfinder} devised a multi-label classifier using machine learning techniques to identify IoT devices, generating a statistical fingerprint for each device using DNS traffic traces. The work in \cite{sentinel} presented a machine-learning-based model to identify the types of IoT devices connecting to a network and enforce security rules in vulnerable ones. The authors in \cite{unsw} suggested a multi-stage machine learning classifier to identify IoT devices based on their network activity. The work in \cite{kolcun2021revisiting} evaluated four machine-learning-based approaches of IoT devices identification, concluding the need to retrain the models to avoid a performance drop. The work in \cite{liu2021} suggested an enhanced deep learning
framework to identify IoT devices. 

Domain name classification using machine learning, on the other hand, is also a recurring focus within research to detect, for example, phishing and DGA-generated domain names. The works in~\cite{nguyet2022, Catal2022} are recent systematic literature reviews about using Deep Learning (DL) techniques for phishing detection. The work in~\cite{durld2021} suggests using DL techniques to detect phishing websites but using raw domain names and applying embedding to each character, conclude that using raw domain names is computationally less expensive than other techniques. The work in~\cite{butnaru2021} also deals with raw URL data in what is referred to as a \emph{lightweight} URL-based phishing detection and uses supervised machine learning techniques to extract features. The authors in~\cite{rao2020} use supervised machine learning to detect phishing URLs based on features extracted from the URL, such as the hostname, full URL, and the Term Frequency-Inverse Document Frequency (TF-IDF), achieving an accuracy of up to 94\%. The authors in~\cite{qiao2019} and~\cite{woodbridge2016} use Long Short-Term Memory (LSTM) networks to detect domain names generated by DGAs. Other techniques include using Generative Adversarial Networks~\cite{geng2022}, full-convolutional systems~\cite{yang2021}, and semi-supervised learning~\cite{faroughi2021}.

%% file: conclusion.tex
\section{Conclusion and Outlook}%
\label{sec:conclusion}
In this paper, we studied the properties of domain names of servers contacted by \emph{IoT~M2M~Devices} and trained several machine learning models to classify between \emph{IoT~M2M~Names} and \emph{Other~Names}. We collected 12 public lists of domain names using past studies and two top lists, \emph{Cisco} and \emph{Tranco}. Our results showed that solely relying on the statistical properties of domain names does not indicate its type. We also observed that the TLDs of \emph{IoT~M2M~Names} are common with \emph{Other~Names} and, therefore, are not indicative of their class. The machine learning models we trained, on the other hand, were successful in classifying \emph{IoT~M2M~Names} and \emph{Other~Names}, with Random Forest having the best (overall) performance. 

Looking forward, we aim to increase the size of the IoT dataset either by adding datasets from future works that have testbeds of real IoT devices or by setting up our own testbed. In addition, we aim to train our models against lists of known malicious domain names and domain names generated by DGAs to enhance the security aspect of our classifier. Future work may also include applying different word embedding techniques and feature extraction methods. %

%% file: appendix.tex
\section{The 12 Datasets Used to Construct the IoT~M2M~Names List}
\label{AppendixA:datasets}
We used 12 datasets from previous works to construct  \emph{IoT~M2M~Names}. The datasets are: IoTFinder~\cite{iotfinder}, YourThings~\cite{yourthings}, MonIoTr~\cite{moniotr}, IoTLS~\cite{iotls}, three datasets from the USC/ISI ANT project~\cite{landerbootup2016, landerbootup2018, landertraces2020}, Edge-IIoTset~\cite{edge}, IoT Sentinel~\cite{sentinel}, IoT Network Intrusion Dataset~\cite{kang}, UNSW IoT traffic traces~\cite{unsw}, and UNSW IoT attack traces~\cite{unsw_attack}. Below is a summary of each one:

\begin{itemize}
    
    \item \textbf{IoTFinder}~\cite{iotfinder}: IoTFinder is a multi-label classifier for detecting IoT devices by studying passively collected DNS traffic. The testbed contained 65 IoT devices. The data were collected between August~1,~2019, and September~30,~2019.
    \item \textbf{YourThings}~\cite{yourthings}: A study of home-based IoT devices to assess their security properties. The testbed contained 65 IoT devices. The data were collected between April~10 and April~19,~2018.
    \item \textbf{MonIoTr}~\cite{moniotr}: A study of information exposed in the traffic of consumer IoT devices. The testbed contained 81 IoT devices. The data were collected between March~28 and May~8, 2019, as well as on September~1,~2019.
    \item \textbf{IoTLS}~\cite{iotls}: A study about the use of TLS in consumer IoT devices. The testbed contained 40 IoT devices. The data were collected between January~2018 and March~2020.
    \item \textbf{USC/ISI ANT project}~\cite{landerbootup2016, landerbootup2018, landertraces2020}: The ANT Lab is an Internet research group at the University of Southern California (USC) that has published several datasets related to various network topics e.g., traffic, outage, and DNS\@. We used three datasets from the USC/ISI ANT project. Two datasets contain the bootup traces of 6 and 11 IoT devices, respectively. The third dataset contains traffic observed in a network of 14 IoT devices over a period of 10 days. 
    \item \textbf{Edge-IIoTset}~\cite{edge}: An IoT traffic dataset that includes benign and attack traffic. The testbed contained 13 real IoT devices. The benign traffic, which we used in this paper, was collected between November~21,~2021, and January~10,~2022. 
    \item \textbf{IoT SENTINEL}~\cite{sentinel}: IoT SENTINEL is a security system that identifies devices present in the network and monitors traffic from vulnerable ones. The testbed contained 31 real IoT devices, and the traffic was collected during the setup of each device.
    \item \textbf{IoT Network Intrusion Dataset}~\cite{kang}: An IoT traffic dataset that includes benign and attack traffic. The testbed contained two real IoT devices. We used the benign traffic in this paper.   
    \item \textbf{UNSW IoT traffic traces}~\cite{unsw}: A study about classification of IoT devices in Smart Home environments. The testbed contained 28 IoT devices, and the traffic was collected between October~2016 and April~2017.
    \item \textbf{UNSW IoT  attack traces}~\cite{unsw_attack}: A study about detecting volumetric attacks against IoT devices and the dataset includes benign and attack traffic. The testbed contained 10 IoT devices, and the traffic was collected for 16 days.
    \end{itemize}

\paragraph{Ethical Considerations} This work does not raise any ethical issues.